\documentclass{PoS}
\usepackage{epsfig,graphicx,bm}
\usepackage{amssymb}
\usepackage{amsmath}

\title{Eta, Eta' and eLSM}

\ShortTitle{Eta, Eta' and eLSM}

\author{\speaker{Denis Parganlija}
\\
Institute for Theoretical Physics, Vienna University of Technology,
Wiedner Hauptstr.\ 8-10, A-1040 Vienna, Austria\\
        E-mail: \email{denisp@hep.itp.tuwien.ac.at}}

\author{P{\'e}ter Kov{\'a}cs\\
        Institute for Particle and Nuclear Physics, Wigner Research Center for
Physics, Hungarian Academy of Sciences, H-1525 Budapest, Hungary\\
        E-mail: \email{kovacs.peter@wigner.mta.hu}}
        
\author{Gy{\"o}rgy Wolf\\
       Institute for Particle and Nuclear Physics, Wigner Research Center for
Physics, Hungarian Academy of Sciences, H-1525 Budapest, Hungary\\
        E-mail: \email{wolf.gyorgy@wigner.mta.hu}}
 
\author{Francesco Giacosa \\
       Institute for Theoretical Physics, Johann Wolfgang Goethe University,
Max-von-Laue-Str.\ 1, D-60438 Frankfurt am Main, Germany\\
        E-mail: \email{giacosa@th.physik.uni-frankfurt.de}}

\author{Dirk H.\ Rischke\\

Institute for Theoretical Physics, Johann Wolfgang Goethe University,
Max-von-Laue-Str.\ 1, D-60438 Frankfurt am Main, Germany and \\
Frankfurt Institute for Advanced Studies, Ruth-Moufang-Str.\ 1 
D-60438
Frankfurt am Main, Germany\\
        E-mail: \email{drischke@th.physik.uni-frankfurt.de}}       

\abstract{We discuss the $\eta$-$\eta^{\prime}$ mixing in the Extended Linear Sigma Model (eLSM).}

\FullConference{Xth Quark Confinement and the Hadron Spectrum,\\
		October 8-12, 2012\\
		TUM Campus Garching, Munich, Germany}

\begin{document}

\section{Introduction}

Quantum Chromodynamics (QCD) -- the fundamental theory of strong interaction -- exhibits several symmetry properties.
Among others, QCD
possesses an exact $SU(3)_{c}$ local gauge symmetry (the colour symmetry) and
an approximate global $U(N_{f})_{R}\times U(N_{f})_{L}$ symmetry for $N_{f}$
massless quark flavours (the chiral symmetry). For sufficiently low
temperature and density quarks and gluons are confined into colourless hadrons
[i.e., $SU(3)_{c}$ invariant configurations]. Thus, in the low-energy region, the chiral symmetry
is the one predominantly determining hadron interactions.

Such interactions are governed not only by the chiral symmetry itself but also by the breaking patterns of this symmetry. 
There are two mechanisms of chiral-symmetry breaking:
explicit, via non-vanishing quark masses, and spontaneous,
via the so-called quark condensate \cite{refssb}. 

The chiral symmetry is isomorphic to the $U(N_{f}%
)_{V}\times U(N_{f})_{A}\equiv U(1)_{V}\times SU(N_{f})_{V}\times
U(1)_{A}\times SU(N_{f})_{A}$ symmetry. Classically, if quark masses 
are non-vanishing but degenerate, then the symmetry is broken explicitly
to $ U(1)_{V}\times SU(N_{f})_{V}$
and for non-degenerate quark masses it is broken to $U(1)_{V}$. 

As already indicated, chiral symmetry is approximate if quark flavours are massless. In fact, the symmetry is even exact in the chiral limit -- but only classically.
At the quantum level, the $U(1)_A$ component of the symmetry is broken spontaneously
by the so-called chiral anomaly \cite{refanomaly}.

Another source of the spontaneous chiral-symmetry breaking is the quark condensate
\begin{equation}
\langle\bar{q}q\rangle=\langle0|\bar{q}q|0\rangle=-i\text{Tr}\lim
_{y\rightarrow x^{+}}S_{F}(x,y) \label{chiralc}
\end{equation}
where $S_{F}(x,y)$ denotes the full quark propagator. The condensate breaks the symmetry $SU(N_{f})_{V}\times SU(N_{f})_{A}$ to $SU(N_{f})_{V}$ with
$N_f^2 - 1$ less generators in the residual than in the original symmetry group.

Consequently, the emergence of $N_{f}^{2}-1$ massless pseudoscalar degrees of freedom (Goldstone bosons) is expected from the Goldstone Theorem \cite{GT}. This is indeed observed: e.g., for
$N_{f}=2$, pions represent long-established lightest degrees of freedom in QCD. Their mass is, however, not zero but rather close to $\sim 140$ MeV due to the explicit breaking of
the chiral symmetry, rendering them pseudo-Goldstone bosons. For $N_{f}=3$,
experimental observations yield five additional pseudoscalar Goldstone states:
four kaons and the $\eta$ meson. However, experimental data also demonstrate the existence of an additional pseudoscalar degree of freedom: $\eta^{\prime}$.

The existence of $\eta^{\prime}$ cannot be explained only by the symmetry-breaking 
pattern $SU(N_{f})_{V}\times SU(N_{f})_{A}$ $\rightarrow SU(N_{f})_{V}$ but rather requires
a broader symmetry-breaking mechanism reading $U(N_{f})_{V}$ $\times U(N_{f})_{A} \rightarrow U(N_{f})_{V}$
where the symmetry corresponding to the one-dimensional
$U(1)_A$ group is also broken, both explicitly and spontaneously. $U(1)_A$ properties are then related to those of $\eta^{\prime}$ -- including the axial anomaly
which implies that $\eta^{\prime}$ remains massless even in the 
chiral limit. (For a review of $\eta$ and $\eta^{\prime}$, see, e.g., 
Ref.\ \cite{Kupsc} and references therein.)

Note, however, that properties of $\eta^{\prime}$ cannot be considered separately from those of the $\eta$ meson: they possess the same quantum numbers and can therefore mix.
It is important to understand the mixing pattern of these two states for at least two reasons: ({\it i}) it allows one to study their structures
(relative contributions of $u$, $d$ and $s$ quarks) -- see below 
-- and, correspondingly, ({\it ii}) it enables us to understand
the decay patterns of $\eta$ and $\eta^{\prime}$, some of which give us insight into the famous CP violation in the pseudoscalar sector \cite{Kupsc}.

The mixing of $\eta$ and $\eta^{\prime}$ mesons can be studied within the realm of the three-flavour Linear Sigma Model with vector and axial-vector degrees of freedom (extended Linear Sigma Model - eLSM). The model contains
only quarkonia, i.e., $\bar q q$ states \cite{Doktorarbeit, Paper1, Paper3}, rendering it suitable to study quarkonium mixing
in various channels including the $\eta$-$\eta^{\prime}$ one. The physical states are obtained from mixing of two pure states: 
$\eta_N \equiv (\bar{u}u + \bar{d}d) / \sqrt{2}$ and $\eta_S \equiv \bar{s}s$. Thus our approach allows us to study quarkonium content,
as well as the mixing angle, of $\eta$ and $\eta^\prime$.

We emphasise, however, that the inclusion of vector and 
axial-vector degrees of freedom into our model is necessary and important since (axial-)vectors 
influence phenomenology in other channels \cite
{Paper1}. Note that glue admixtures to $\eta$ and $\eta^{\prime}$
can also be considered along the lines of Ref.\ \cite{Stani} but
that will not be of concern in the present article.

This paper is organised as follows. In Sec.\ \ref{L} we discuss the eLSM Lagrangian; in Sec.\ \ref{sec.eta-eta} we describe the $\eta$-$\eta^\prime$ mixing pattern
and in Sec.\ \ref{C} we present a conclusion and an outlook of our work.

\section{The Lagrangian} \label{L}

The globally invariant $U(3)_{L}\times U(3)_{R}$ Lagrangian possesses the following structure \cite{Doktorarbeit,Paper1,Paper3,References,References2,Peniche}:
\begin{align}
\mathcal{L}  &  =\mathrm{Tr}[(D_{\mu}\Phi)^{\dagger}(D_{\mu}\Phi)]-m_{0}%
^{2}\mathrm{Tr}(\Phi^{\dagger}\Phi)-\lambda_{1}[\mathrm{Tr}(\Phi^{\dagger}%
\Phi)]^{2}-\lambda_{2}\mathrm{Tr}(\Phi^{\dagger}\Phi)^{2}\nonumber\\
&  -\frac{1}{4}\mathrm{Tr}(L_{\mu\nu}^{2}+R_{\mu\nu}^{2})+\mathrm{Tr}\left[
\left(  \frac{m_{1}^{2}}{2}+\Delta\right)  (L_{\mu}^{2}+R_{\mu}^{2})\right]
+\mathrm{Tr}[H(\Phi+\Phi^{\dagger})]\nonumber\\
&  +c_{1}(\det\Phi-\det\Phi^{\dagger})^{2}+i\frac{g_{2}}{2}(\mathrm{Tr}%
\{L_{\mu\nu}[L^{\mu},L^{\nu}]\}+\mathrm{Tr}\{R_{\mu\nu}[R^{\mu},R^{\nu
}]\})\nonumber\\
&  +\frac{h_{1}}{2}\mathrm{Tr}(\Phi^{\dagger}\Phi)\mathrm{Tr}(L_{\mu}%
^{2}+R_{\mu}^{2})+h_{2}\mathrm{Tr}[|L_{\mu}\Phi|^{2}+|\Phi R_{\mu}%
|^{2}]+2h_{3}\mathrm{Tr}(L_{\mu}\Phi R^{\mu}\Phi^{\dagger})\text{.}
\label{Lagrangian}%
\end{align}

The scalar and pseudoscalar states present in Eq.\ (\ref{Lagrangian}) are:

\begin{equation}
\Phi=\frac{1}{\sqrt{2}}\left(
\begin{array}
[c]{ccc}%
\frac{(\sigma_{N}+a_{0}^{0})+i(\eta_{N}+\pi^{0})}{\sqrt{2}} & a_{0}^{+}%
+i\pi^{+} & K_{0}^{\star +}+iK^{+}\\
a_{0}^{-}+i\pi^{-} & \frac{(\sigma_{N}-a_{0}^{0})+i(\eta_{N}-\pi^{0})}%
{\sqrt{2}} & K_{0}^{\star 0}+iK^{0}\\
K_{0}^{\star -}+iK^{-} & {\bar{K}_{0}^{\star 0}}+i{\bar{K}^{0}} & \sigma_{S}+i\eta_{S}%
\end{array}
\right)
\end{equation}

The matrices containing vectors and axial-vectors read:

\begin{equation}
V^{\mu}=\frac{1}{\sqrt{2}}\left(
\begin{array}
[c]{ccc}%
\frac{\omega_{N}^{\mu}+\rho^{\mu0}}{\sqrt{2}} & \rho^{\mu+} & K^{\star\mu+}\\
\rho^{\mu-} & \frac{\omega_{N}^{\mu}-\rho^{\mu0}}{\sqrt{2}} & K^{\star\mu0}\\
K^{\star\mu-} & {\bar{K}}^{\star\mu0} & \omega_{S}^{\mu}%
\end{array}
\right), \qquad 
A^\mu=\frac{1}{\sqrt{2}}\left(
\begin{array}
[c]{ccc}%
\frac{f_{1N}^{\mu}+a_{1}^{\mu0}}{\sqrt{2}} & a_{1}^{\mu+} & K_{1}^{\mu+}\\
a_{1}^{\mu-} & \frac{f_{1N}^{\mu}-a_{1}^{\mu0}}{\sqrt{2}} & K_{1}^{\mu0}\\
K_{1}^{\mu-} & {\bar{K}}_{1}^{\mu0} & f_{1S}^{\mu}%
\end{array}
\right)   \label{Ve}%
\end{equation}
with the right-handed (axial-)vector matrix $R^\mu = V^\mu - 
A^\mu$ and the left-handed (axial-)vector matrix $L^\mu = V^\mu 
- A^\mu$. Additionally, $D^{\mu}\Phi = \partial^{\mu}\Phi-ig_{1}(L^{\mu}\Phi-\Phi R^{\mu
})$ $-ie{\tilde A}^{\mu}[T_{3},\Phi]$ is the covariant
derivative; $L^{\mu \nu }=\partial ^{\mu }L^{\nu }-ie{\tilde A}^{\mu }[T_{3},L^{\nu
}]-\{ \partial ^{\nu }L^{\mu }$ $-ie{\tilde A}^{\nu }[T_{3},L^{\mu }] \} $ and $R^{\mu \nu }=\partial ^{\mu }R^{\nu }-ie{\tilde A}^{\mu }[T_{3},
R^{\nu }]-\left\{ \partial ^{\nu
}R^{\mu }-ie{\tilde A}^{\nu }[T_{3},R^{\mu }]\right\} $ are, respectively, the
left-handed and right-handed field strength tensors, ${\tilde A}^{\mu }$ is the
electromagnetic field, $T_{3}$ is the third generator of the $SU(3)$ group
and the term $c_{1}(\det \Phi -\det \Phi ^{\dagger })^{2}$ describes the $%
U(1)_{A}$ anomaly \cite{Klempt}.

Explicit breaking of the global symmetry in the (pseudo)scalar channel is
described by the term Tr$[H(\Phi+\Phi^{\dagger})]$ and
in the (axial-)vector channel by the term $\mathrm{Tr}\left[ \Delta(L_{\mu
}^{2}+R_{\mu}^{2})\right]$ where $H= \text{diag}(h_{0N},h_{0N},h_{0S})$ and $\Delta = \text{diag}(\delta_{N}, \delta_{N}, \delta_{S})$
with $h_N \sim m_u = m_d$, $h_S \sim m_s$, $\delta_N \sim m_u^2$ and $\delta_S \sim m_s^2$ (isospin symmetry assumed in the non-strange sector).  

We assign the field $\vec{\pi}$ to the pion; $\eta _{N}$ and $\eta_S$ are assigned, respectively, to the pure non-strange 
and the pure strange counterparts of the $\eta$ and $\eta^\prime$ mesons. The fields $\omega _{N}^{\mu }$, $\vec{\rho}^{\mu }$, $f_{1N}^{\mu }$
and $\vec{a}_{1}^{\mu }$ are assigned to the $\omega (782)$, $\rho (770)$, $%
f_{1}(1285)$ and $a_{1}(1260)$ mesons, respectively. We also
assign the $K$ fields to the kaons; the $\omega _{S}^{\mu }$, $%
f_{1S}^{\mu }$ and $K^{\star \mu }$ fields correspond to the $\varphi (1020)$%
, $f_{1}(1420)$ and $K^{\star }(892)$ mesons, respectively. Assignment of
the $K_{1}^{\mu }$ field is, unfortunately, not as clear since this state
can be assigned either to the $K_{1}(1270)$ or to the $K_{1}(1400)$
resonances \cite{Peniche} but that is of no importance for the following. 

Spontaneous breaking of the chiral symmetry
is implemented by considering the respective non-vanishing vacuum expectation values $\phi_N$ and $\phi_S$ of
the two scalar isosinglet states present in our model, $\sigma_{N}%
\equiv(\bar{u}u+\bar{d}d)/\sqrt{2}$ and 
$\sigma_{N}\equiv\bar{s}s$. The
relations between $\phi_{N,S}$ and the pion decay constant 
$f_{\pi}$ as well
as the kaon decay constant $f_{K}$ read
$f_{\pi}=\phi_{N}/Z_{\pi}$ and $f_{K}=$ $\left(  \sqrt{2} \phi_{S} + \phi
_{N}\right)  /(2 Z_{K})$
where $f_{\pi}=92.4$ MeV and $f_{K}=155.5/\sqrt{2}$ MeV \cite{PDG}. The
chiral condensates $\phi_{N}$ and $\phi_{S}$ lead to mixing
terms in the Lagrangian (\ref{Lagrangian}) that need to be removed in order for the scattering matrix stemming from the Lagrangian
to be diagonal (the procedure is described detailedly in Ref.\ \cite{Doktorarbeit}).

\section{Mixing of \boldmath $\eta$ and \boldmath $\eta^{\prime}$} \label{sec.eta-eta}

Our Lagrangian (\ref{Lagrangian}) yields the following mixing term of the pure non-strange and strange fields $\eta_{N}$ and $\eta_{S}$:
\begin{equation}
\mathcal{L}_{\eta_{N}\eta_{S}}=-c_{1}\frac{Z_{\eta_{S}}Z_{\pi}}{2}\phi_{N}%
^{3}\phi_{S}\eta_{N}\eta_{S}\text{.} \label{eta-eta}%
\end{equation}

The full $\eta_{N}$-$\eta_{S}$\ interaction Lagrangian obtained from Eq.\ (\ref{Lagrangian}) has the form%
\begin{equation}
\mathcal{L}_{\eta_{N}\eta_{S,}\,\mathrm{full}}=\frac{1}{2}(\partial_{\mu}%
\eta_{N})^{2}+\frac{1}{2}(\partial_{\mu}\eta_{S})^{2}-\frac{1}{2}m_{\eta_{N}%
}^{2}\eta_{N}{}^{2}-\frac{1}{2}m_{\eta_{S}}^{2}\eta_{S}{}^{2}+z_{\eta}\eta
_{N}\eta_{S}\text{,} \label{eta-eta-1}%
\end{equation}

where $z_{\eta}$ is the mixing term of the pure states $\eta_{N}\equiv(\bar
{u}u-\bar{d}d)/\sqrt{2}$ and $\eta_{S}\equiv\bar{s}s$:
\begin{equation}
z_{\eta}=-c_{1}\frac{Z_{\eta_{S}}Z_{\pi}}{2}\phi_{N}^{3}\phi_{S}\text{.}
\label{z1}%
\end{equation}

However, mixing between $\eta_{N}$ and $\eta_{S}$ can be
equivalently expressed as the mixing between the octet state%
\begin{equation}
\eta_{8}=\sqrt{\frac{1}{6}}(\bar{u}u+\bar{d}d-2\bar{s}s)\equiv\sqrt{\frac
{1}{3}}\eta_{N}-\sqrt{\frac{2}{3}}\eta_{S} \label{eta_8}%
\end{equation}

and the singlet state%

\begin{equation}
\eta_{0}=\sqrt{\frac{1}{3}}(\bar{u}u+\bar{d}d+\bar{s}s)\equiv\sqrt{\frac{2}%
{3}}\eta_{N}+\sqrt{\frac{1}{3}}\eta_{S}\text{.} \label{eta_0}%
\end{equation}

We determine the physical states $\eta$ and $\eta^{\prime}$ as mixture
of the octet and singlet states with a mixing angle $\varphi_{P}$ (see, e.g., Ref.\ \cite{Feldmann} and refs.\ therein):%
\begin{equation}
\left(
\begin{array}
[c]{c}%
\eta\\
\eta^{\prime}%
\end{array}
\right)  =\left(
\begin{array}
[c]{cc}%
\cos\varphi_{P} & -\sin\varphi_{P}\\
\sin\varphi_{P} & \cos\varphi_{P}%
\end{array}
\right)  \left(
\begin{array}
[c]{c}%
\eta_{8}\\
\eta_{0}%
\end{array}
\right) \label{aaa}
\end{equation}

or, using Eqs.\ (\ref{eta_8}) and (\ref{eta_0}),

\begin{equation}
\left(
\begin{array}
[c]{c}%
\eta\\
\eta^{\prime}%
\end{array}
\right)  =\left(
\begin{array}
[c]{cc}%
\cos\varphi_{P} & -\sin\varphi_{P}\\
\sin\varphi_{P} & \cos\varphi_{P}%
\end{array}
\right)  \left(
\begin{array}
[c]{cc}%
\sqrt{\frac{1}{3}} & -\sqrt{\frac{2}{3}}\\
\sqrt{\frac{2}{3}} & \sqrt{\frac{1}{3}}%
\end{array}
\right)  \left(
\begin{array}
[c]{c}%
\eta_{8}\\
\eta_{0}%
\end{array}
\right)  \text{.}%
\end{equation}

If we introduce $\arcsin(\sqrt{2/3})=54.7456%
{{}^\circ}%
\equiv\varphi_{I}$, then the trigonometric addition formulas lead to

\begin{equation}
\left(
\begin{array}
[c]{c}%
\eta\\
\eta^{\prime}%
\end{array}
\right)  =\left(
\begin{array}
[c]{cc}%
\cos(\varphi_{P}+\varphi_{I}) & -\sin(\varphi_{P}+\varphi_{I})\\
\sin(\varphi_{P}+\varphi_{I}) & \cos(\varphi_{P}+\varphi_{I})
\end{array}
\right)  \left(
\begin{array}
[c]{c}%
\eta_{N}\\
\eta_{S}%
\end{array}
\right)  \text{.}%
\end{equation}

Defining the $\eta$-$\eta^{\prime}$ mixing angle $\varphi_{\eta}$
\begin{equation}
\varphi_{\eta}=-(\varphi_{P}+\varphi_{I})\text{,} \label{phi_eta}%
\end{equation}

we obtain

\begin{equation}
\left(
\begin{array}
[c]{c}%
\eta\\
\eta^{\prime}%
\end{array}
\right)  =\left(
\begin{array}
[c]{cc}%
\cos\varphi_{\eta} & \sin\varphi_{\eta}\\
-\sin\varphi_{\eta} & \cos\varphi_{\eta}%
\end{array}
\right)  \left(
\begin{array}
[c]{c}%
\eta_{N}\\
\eta_{S}%
\end{array}
\right)
\end{equation}

or in other words
\begin{align}
\eta &  =\cos\varphi_{\eta}\eta_{N}+\sin\varphi_{\eta}\eta_{S}\text{,} \label{eta}\\
\eta^{\prime}  &  =-\sin\varphi_{\eta}\eta_{N}+\cos\varphi_{\eta}\eta
_{S}\text{.} \label{etap}%
\end{align}

The Lagrangian in Eq.\ (\ref{eta-eta-1}) contains only pure states
$\eta_{N}$ and $\eta_{S}$. Inverting Eqs.\ (\ref{eta}) and (\ref{etap})
\begin{align}
\eta_{N}  &  =\cos\varphi_{\eta}\eta-\sin\varphi_{\eta}\eta^{\prime
}\text{,} \label{etaN}\\
\eta_{S}  &  =\sin\varphi_{\eta}\eta+\cos\varphi_{\eta}\eta^{\prime}\text{,}
\label{etaS}%
\end{align}

we can isolate the relevant part of the Lagrangian (\ref{Lagrangian}) and determine the parametric form of the $\eta$
and $\eta^\prime$ mass terms that read
\begin{align}
m_{\eta}^{2}  &  =m_{\eta_{N}}^{2}\cos^{2}\varphi_{\eta}+m_{\eta_{S}}^{2}%
\sin^{2}\varphi_{\eta}-z_{\eta}\sin(2\varphi_{\eta})\text{,} \label{m_eta}\\
m_{\eta^{\prime}}^{2}  &  =m_{\eta_{N}}^{2}\sin^{2}\varphi_{\eta}+m_{\eta_{S}%
}^{2}\cos^{2}\varphi_{\eta}+z_{\eta}\sin(2\varphi_{\eta})
\label{m_eta'}%
\end{align}

where
\begin{eqnarray}
m_{\eta_{N}}^{2}  &  =Z_{\pi}^{2}\left[  m_{0}^{2}+\left(  \lambda_{1}%
+\frac{\lambda_{2}}{2}\right)  \phi_{N}^{2}+\lambda_{1}\phi_{S}^{2}+c_{1}%
\phi_{N}^{2}\phi_{S}^{2}\right] \text{,} \label{m_eta_N} \\
m_{\eta_{S}}^{2}  &  =Z_{\eta_{S}}^{2}\left[  m_{0}^{2}+\lambda_{1}\phi
_{N}^{2}+(\lambda_{1}+\lambda_{2})\phi_{S}^{2}+c_{1}\frac{\phi_{N}^{4}}%
{4}\right] \label{m_eta_S} 
\end{eqnarray}
are the parametric forms of the pure states $\eta_N$ and $\eta_S$. [Note that the determination of $m_{\eta_{N}}$ and $m_{\eta_{N}}$ and, consequently,
$m_{\eta}$, $m_{\eta^\prime}$ and $\varphi_\eta$ requires a fit of {\it all} parameters in the Lagrangian (\ref{Lagrangian}) that, in turn, requires much more observables
than the above two isosinglet masses. The fit procedure will be described further below.]

Assigning our fields $\eta$ and $\eta^{\prime}$ to physical (asymptotic) states requires that
the Lagrangian $\mathcal{L}_{\eta\eta^{\prime}}$ does not contain
any $\eta$-$\eta^{\prime}$ mixing terms leading to the condition
\begin{equation}
z_{\eta}\overset{!}{=}(m_{\eta_{S}}^{2}-m_{\eta_{N}}^{2})\tan(2\varphi_{\eta
})/2\text{.} \label{z}%
\end{equation}

\subsection{Fit Procedure}

In order to make statements regarding $\eta$-$\eta^\prime$ mixing we first need to determine values of our model parameters.
Lagrangian (\ref{Lagrangian}) contains 14 parameters: $\lambda _{1}$, $%
\lambda _{2}$, $c_{1}$, $h_{0N}$, $h_{0S}$, $h_{1}$, $h_{2}$, $h_{3}$, $%
m_{0}^{2}$, $g_{1}$, $g_{2}$, $m_{1}$, $\delta _{N}$, $\delta _{S}$.
Parameters $h_{0N}$ and $h_{0S}$ are determined from the extremum condition
for the potential obtained from Eq.\ (\ref{Lagrangian}). Parameter $\delta
_{N}$ is set to zero throughout this paper because the explicit
breaking of the chiral symmetry is small in the non-strange quark sector. The other 11 parameters
are calculated from a global fit including 21 observables \cite{Paper3}: $f_{\pi }$, $%
f_{K} $, $m_{\pi }$, $m_{K}$, $m_{\eta }$, $m_{\eta ^{\prime }}$, $m_{\rho }$%
, $m_{K^{\star }}$, $m_{\omega _{S}\equiv \varphi (1020)}$, $m_{f_{1S}\equiv
f_{1}(1420)}$, $m_{a_{1}}$, $m_{a_{0}\equiv a_{0}(1450)}$, $m_{K_{0}^{\star
}\equiv K_{0}^{\star }(1430)}$, $\Gamma _{\rho \rightarrow \pi \pi }$, $%
\Gamma _{K^{\star }\rightarrow K\pi }$, $\Gamma _{\phi \rightarrow KK}$, $%
\Gamma _{a_{1}\rightarrow \rho \pi }$, $\Gamma _{a_{1}\rightarrow \pi \gamma
}$, $\Gamma _{f_{1}(1420)\rightarrow K^{\star }K}$, $\Gamma _{a_{0}(1450)}$, 
$\Gamma _{K_{0}^{\star }(1430)\rightarrow K\pi }$ 
(data from PDG \cite{PDG}; since our model currently implements no isospin-symmetry breaking that would influence the physical hadron masses by the order of 5\%,
we have modified error values presented by PDG
such that we use the PDG error value if it is larger than 5\% and increase the error to 5\% otherwise.)
 Note that the observables entering the fit allow us to determine only
linear combinations $m_{0}^{2}+\lambda _{1}(\phi _{N}^{2}+\phi _{S}^{2})$
and $m_{1}^{2}+h_{1}\left( \phi _{N}^{2}+\phi _{S}^{2}\right) /2$ rather
than parameters $m_{0}$, $m_{1}$, $\lambda _{1}$ and $h_{1}$ by themselves.
However, it is nonetheless possible to calculate masses of $\eta$ and $\eta^\prime$ as evident from Eqs.\ (\ref{m_eta}) -- (\ref{m_eta_S}).

\begin{table}[h]
\centering
\begin{tabular}
[c]{|c|c|c|}\hline
Observable & Fit [MeV] & Experiment [MeV]\\\hline
$f_{\pi}$ & $96.3 \pm0.7 $ & $92.2 \pm4.6$\\\hline
$f_{K}$ & $106.9 \pm0.6$ & $110.4 \pm5.5$\\\hline
$m_{\pi}$ & $141.0 \pm5.8$ & $137.3 \pm6.9$\\\hline
$m_{K}$ & $485.6 \pm3.0$ & $495.6 \pm24.8$\\\hline
$m_{\eta}$ & $509.4 \pm3.0$ & $547.9 \pm27.4$\\\hline
$m_{\eta^{\prime}}$ & $962.5 \pm5.6$ & $957.8 \pm47.9$\\\hline
$m_{\rho}$ & $783.1 \pm7.0$ & $775.5 \pm38.8$\\\hline
$m_{K^{\star}}$ & $885.1 \pm6.3$ & $893.8 \pm44.7$\\\hline
$m_{\phi}$ & $975.1 \pm6.4$ & $1019.5 \pm51.0$\\\hline
$m_{a_{1}}$ & $1186 \pm6$ & $1230 \pm62$\\\hline
$m_{f_{1}(1420)}$ & $1372.5 \pm5.3$ & $1426.4 \pm71.3$\\\hline
$m_{a_{0}}$ & $1363 \pm1$ & $1474 \pm74$\\\hline
$m_{K_{0}^{\star}}$ & $1450 \pm1$ & $1425 \pm71$\\\hline
$\Gamma_{\rho\rightarrow\pi\pi}$ & $160.9 \pm4.4$ & $149.1 \pm7.4$\\\hline
$\Gamma_{K^{\star}\rightarrow K\pi}$ & $44.6 \pm1.9$ & $46.2 \pm2.3$\\\hline
$\Gamma_{\phi\rightarrow\bar{K}K}$ & $3.34 \pm0.14$ & $3.54 \pm0.18$\\\hline
$\Gamma_{a_{1}\rightarrow\rho\pi}$ & $549 \pm43$ & $425 \pm175$\\\hline
$\Gamma_{a_{1}\rightarrow\pi\gamma}$ & $0.66 \pm0.01$ & $0.64 \pm0.25$\\\hline
$\Gamma_{f_{1}(1420)\rightarrow K^{\star}K}$ & $44.6 \pm39.9$ & $43.9 \pm
2.2$\\\hline
$\Gamma_{a_{0}}$ & $266 \pm12$ & $265 \pm13$\\\hline
$\Gamma_{K_{0}^{\star}\rightarrow K\pi}$ & $285 \pm12$ & $270 \pm80$\\\hline
\end{tabular}
\caption{Best-fit results for masses and decay widths compared with
experiment.}%
\label{Comparison1}%
\end{table}

In Table \ref{Comparison1} we present our best-fit results (that also include values of $m_{\eta}$ and $m_{\eta^\prime}$).
One of the conditions entering our fit was $m_{\eta_N} < 
m_{\eta_S}$, i.e., pure non-strange states should be lighter 
than pure strange states (for details, see Ref.\ \cite{Paper3}). 
As evident from Table \ref{Comparison1}, under the constraint 
$m_{\eta_N} < m_{\eta_S}$ our fit optimises 
at $m_{\eta} = (509.4 \pm 3.0)$ MeV, below the experimentally determined interval
(presented in the third column of Fig.\ \ref{Comparison1}).
However, the value $m_{\eta^\prime} = (962.5 \pm 5.6)$ MeV is within the experimental boundaries. Note that
the stated values of $m_{\eta}$ and $m_{\eta^\prime}$ also imply $m_{\eta_N} \simeq 766$ MeV and $m_{\eta_N} \simeq 770$ MeV
-- the non-strange and strange mass contributions to $\eta$ and $\eta^\prime$ thus appear to be virtually equal. Furthermore, the mentioned non-saturation
of physical $\eta$ and $\eta^\prime$ masses in our quarkonium-based approach may hint to additional contributions to $\eta$/$\eta^\prime$ spectroscopic
wave functions that go beyond
the antiquark-quark structure.

Our fit determined all parameter values uniquely and therefore the $\eta$-$\eta^\prime$ mixing angle is also uniquely determined as
\begin{equation}
\varphi_\eta = -44.6^{\circ}.
\end{equation}
 
The result is close to maximal mixing, i.e., our result
  suggests a slightly larger mixing than results of Ref.\ \cite{etaM}.
  
\section{Conclusions} \label{C}

We have presented an extended Linear Sigma Model containing vector and axial-vector degrees of freedom (eLSM). A global fit 
of masses and decay widths has been performed in order to study, among others, mixing of isosinglet states in
the pseudoscalar channel ($\eta$-$\eta^\prime$ mixing). The model presented in this article contains no free parameters -- all 
parameters are fixed from the mentioned global fit. In the $\eta$-$\eta^\prime$ channel, our fit optimises at $m_{\eta^\prime} = (962.5 \pm 5.6) $ MeV and 
$m_{\eta} = (509.4 \pm 3.0)$ MeV: 
the former is exactly within the data interval but the latter is below the experimental result.
This may represent a hint of non-$\bar{q}q$ contributions to the pseudoscalar isosinglets.
We have also determined the $\eta$-$\eta^\prime$ mixing angle to be
$\varphi_\eta = -44.6^{\circ}$, close to maximal mixing.

We emphasise, however, that the stated results present only a small part of meson phenomenology that can be
considered in eLSM: the model can also be utilised to study the structure of scalar and axial-vector mesons 
\cite{Doktorarbeit, Paper3, References2, Peniche} but also extended to finite temperatures and densities, similar to Ref.\ \cite{Struber}.

\end{document}